\documentclass[]{bytedance_seed}

\usepackage[toc,page,header]{appendix}

\usepackage{minitoc}
\usepackage{amsmath,amssymb}
\usepackage{bm}
\usepackage{wrapfig}

\usepackage{enumitem}

\title{ Seed3D 2.0: Advancing High-Fidelity Simulation-Ready 3D Content Generation}

\affiliation[]{ByteDance Seed}

\contribution{See \hyperref[sec:contributions]{Contributions and Acknowledgments} section for a full author list.}

\abstract{
We present Seed3D 2.0, an advanced 3D content generation 
system built on Seed3D 1.0~\cite{seed2025seed3d}, with 
substantial improvements across generation fidelity, 
simulation-ready capabilities, and application coverage. 
For geometry, a coarse-to-fine two-stage pipeline decouples global structure learning from high-frequency detail recovery, while a locality-aware VAE achieves higher spatial compression and more efficient decoding. 
For texture and 
material generation, we replace the cascaded pipeline of 
Seed3D 1.0 with a unified PBR model that directly generates 
multi-view albedo and metallic-roughness maps, enhanced by 
Mixture-of-Experts scaling and VLM-based semantic 
conditioning for improved material precision and visual 
fidelity. Beyond single-object generation, Seed3D 2.0 
introduces a simulation-ready model suite comprising scene 
layout planning, part-aware decomposition, and 
training-free articulation generation, enabling coherent 
scene construction and part-level physical interaction 
across physics and graphics engines. 
{A large-scale human preference study against five recent commercial models shows that Seed3D 2.0 achieves consistent win rates of 69.0\% to 89.9\% in 
textured 3D asset generation.}
Seed3D 2.0 is available at \href{https://exp.volcengine.com/ark/vision?_vtm_=0.0.c70961.d701978.0&mode=vision&modelId=doubao-seed3d-2-0-260328&tab=Gen3D}{Volcano Engine}\footnote{Model ID: doubao-seed3d-2-0-260328}.
\vspace{-0.1cm}
}

\checkdata[Official Page]{\url{https://seed.bytedance.com/seed3d_2_0}}
\correspondence{Jianfeng Zhang (\email{jianfengzhang@bytedance.com})}

\begin{document}
\maketitle
\vspace{-1.1cm}
\begin{figure}[h]
  \centering
  \includegraphics[width=0.95\linewidth]{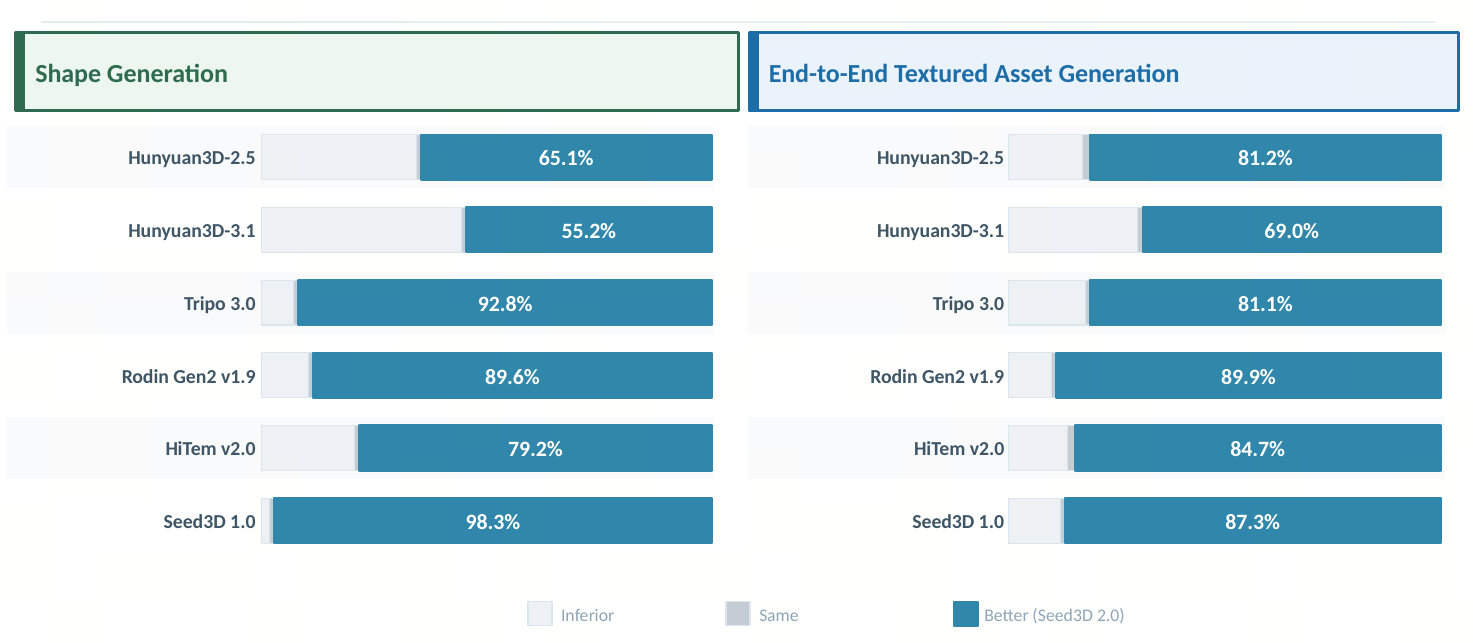}
  \caption{{The results of user study.} {We compare Seed3D 2.0 with the latest commercial 3D generation models on shape-only generation and end-to-end textured asset generation.}}
  \label{fig:quantitative_results}
\end{figure}

\newpage

\tableofcontents
\newpage


\section{Introduction}

The demand for high-quality 3D assets spans an increasingly broad range of applications, including XR content creation, 3D printing and robotic simulation, each placing distinct requirements on {3D assets in terms of} geometric precision, material fidelity, deployment compatibility with physics and graphics engines, and the support for complex asset structures and interactions. Seed3D 1.0~\cite{seed2025seed3d} established a strong foundation for simulation-ready 3D asset generation from single images, but as deployment demands have deepened across these applications, two gaps have become increasingly apparent.

The first is a \textit{quality gap}. Production environments require geometric regularity, including sharp edges and structurally consistent surfaces, alongside physically-based materials that remain visually consistent under varied illumination. Existing models, including Seed3D 1.0, produce plausible shapes but consistently fall short on these fine-grained criteria, limiting their utility across high-fidelity rendering and physics simulation pipelines.

The second is a \textit{capability gap}. As applications grow more demanding, static holistic meshes are no longer sufficient: 
{downstream use cases increasingly call for coherent multi-object scenes, functionally decomposed objects, and assets that support part-level physical interaction.}
Seed3D 1.0 offered limited scene generation capabilities and no part-level decomposition or articulation, leaving a substantial gap between generated outputs and downstream application requirements.

\begin{figure}[t]
    \centering
    \includegraphics[width=\linewidth]{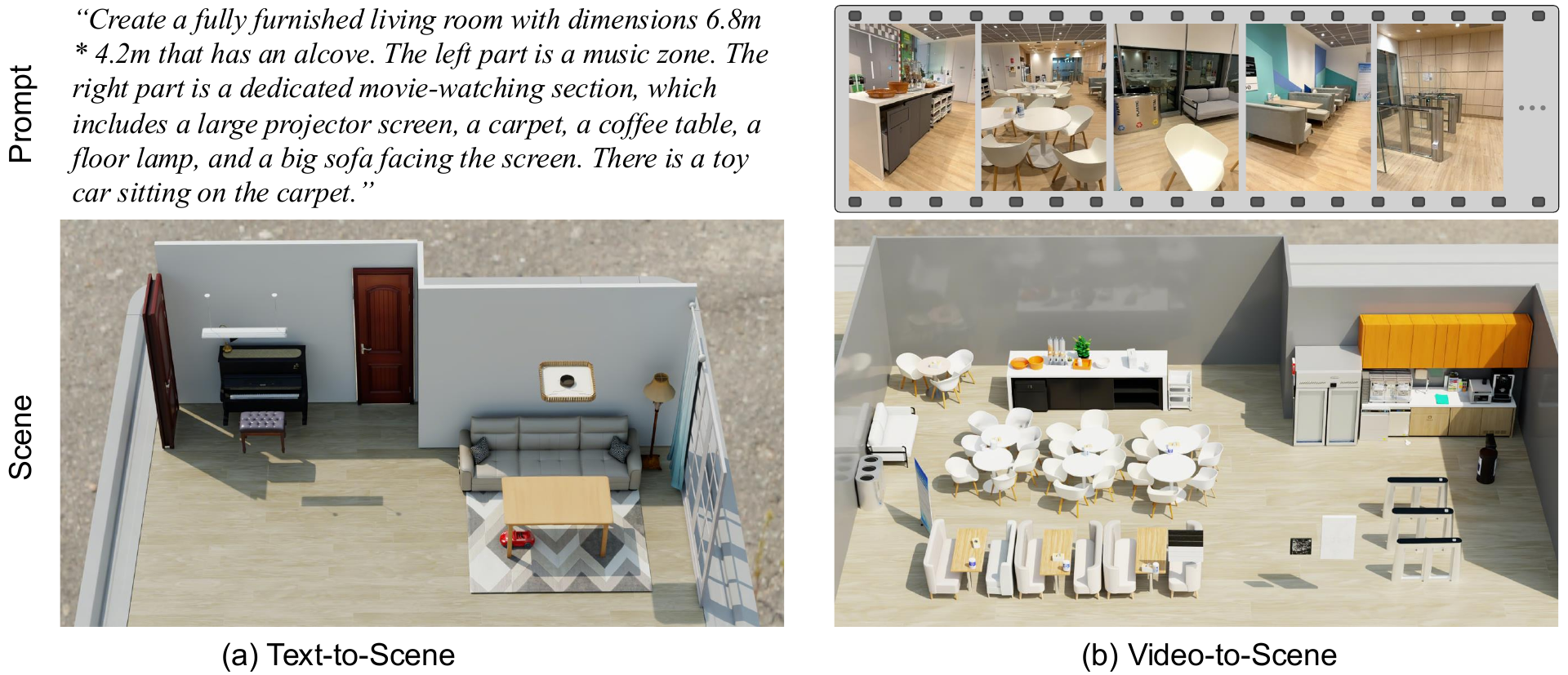}
    \vspace{-6mm}
    \caption{\text{Object-compositional scene generation with Seed3D 2.0.} Given user prompts such as videos or textual descriptions, our scene-layout planning module predicts the 3D spatial layout and then arranges the generated 3D assets into an object-compositional scene. The resulting instance-level 3D scene supports flexible downstream interactions.}
    \label{fig:scene-demo}
\end{figure}

Seed3D 2.0 is designed to systematically close both gaps 
across generation quality and simulation-ready capabilities: 

\begin{itemize}
    \item \textbf{Enhanced Geometry Generation.} Seed3D 1.0 relies on a single-stage generation pipeline, which faces an inherent tension between learning global structure and recovering high-frequency geometric detail. We address this with a coarse-to-fine two-stage generation pipeline: the first stage establishes a geometric foundation, while the second stage specializes in recovering sharp edges and fine surface details conditioned on {global} shape priors and voxelized positional encodings. A locality-aware VAE further improves spatial compression and decoding efficiency, contributing to sharper geometry and higher reconstruction fidelity.

    \item \textbf{Unified PBR Texture Generation.} Rather than the cascaded multi-view RGB synthesis and material estimation pipeline of Seed3D 1.0, we introduce a unified model that directly generates multi-view albedo and metallic-roughness maps from reference images and 3D geometry, eliminating intermediate error accumulation. A Mixture-of-Experts architecture enables higher-resolution generation without proportional computational overhead, while VLM-based semantic conditioning resolves the ill-posed nature of material estimation under unknown illumination. 
    {These advances notably improve 
    material accuracy, texture detail, and fine-grained feature recovery including text rendering fidelity.}

    \item \textbf{Expanded Simulation-Ready Capabilities.} 
    Beyond single-object generation, Seed3D 2.0 introduces three modules to address the capability gap. Scene layout planning supports coherent multi-object scene construction through an automated compositional pipeline (Figure~\ref{fig:scene-demo}). Part-aware generation decomposes assets into functionally meaningful components, enabling fine-grained asset control for downstream applications (Figure~\ref{fig:partgen-demo}). Building on this decomposition, articulation generation infers kinematic structure, joint types, and motion ranges in a training-free manner, producing assets that support part-level physical interaction in physics and graphics engines (Figure~\ref{fig:articulation-demo}).
\end{itemize}

Beyond these core contributions, we further introduce a progressive distillation pipeline that substantially reduces inference cost while preserving generation quality, supporting production-scale deployment. We validate Seed3D 2.0 through a human preference study comparing against five recent commercial models. As shown in Figure~\ref{fig:quantitative_results}, Seed3D 2.0 is consistently preferred over all baselines in end-to-end textured asset generation, with human preference win rates ranging from 69.0\% to 89.9\%.

\newpage
\section{Model Design}

\subsection{Geometry}
Our geometry generation pipeline follows the well-adopted vector-set (VecSet) paradigm~\cite{zhang20233dshape2vecset,hunyuan3d-2.0,li2025triposg} and consists of a 3D variational autoencoder (VAE) and a rectified flow-based diffusion transformer (DiT).

\subsubsection{Seed3D-VAE}
Following the design of Seed3D 1.0~\cite{seed2025seed3d}, Seed3D 2.0 VAE adopts a dual-branch perceiver-based encoder--decoder architecture~\cite{chen2025dora} to compress continuous 3D geometry into a compact VecSet representation composed of latent tokens.
The encoder processes surface point clouds augmented with positional, normal, and sharp-edge samples, mapping them into latent tokens that encode both global topology and fine-grained geometric structure.
The decoder reconstructs a continuous Signed Distance Field (SDF) by attending to the latent tokens via cross-attention over spatial query points, and the final mesh is extracted using Dual Marching Cubes (DMC)~\cite{schaefer2002dual}.

{
To achieve a higher spatial compression ratio, we introduce
\textit{locality-aware latent aggregation}, which exploits the
structural property of VecSet representations~\cite{autopartgen,lattice,lai2025flashvdm}: tokens within the
same spatial neighborhood encode redundant geometric information.
During encoding, we consolidate tokens across spatial neighborhoods,
concentrating representational capacity in geometrically complex
regions while preserving a compact global structure. This design achieves higher reconstruction quality than
Seed3D 1.0 VAE with fewer latent tokens.
The same locality principle extends to \textit{decoding efficiency}.
In SDF decoding, each spatial query must attend to all latent
tokens~\cite{lai2025flashvdm}, making this cross-attention the
dominant computational cost. We replace this dense attention with
a content-adaptive sparse routing mechanism that restricts each
spatial query to a compact, spatially coherent token subset, substantially reducing decoding latency while preserving reconstruction fidelity.
}

\begin{figure}[t]
    \centering
    \includegraphics[width=\linewidth]{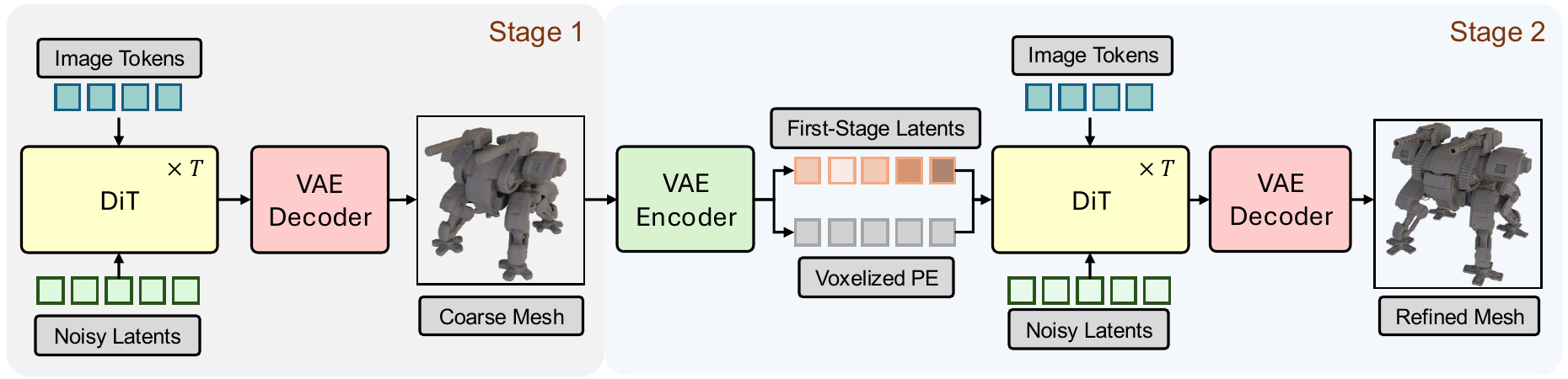}
    \caption{{Overview of the Seed3D 2.0 geometry generation pipeline.} Stage 1 learns to generate coarse geometric structure, while Stage 2 refines the coarse geometry and produces high-frequency details.}
    \label{fig:shape_pipeline}
\end{figure}

\subsubsection{Seed3D-DiT}

Building upon the geometry-aware latent space of Seed3D 2.0 VAE, 
Our DiT employs a rectified flow-based diffusion framework~\cite{lipman2022flow} to 
model the transformation from Gaussian noise to structured latent 
representations. Rather than generating all geometric detail in a 
single pass, we introduce a \textit{coarse-to-fine two-stage 
generation strategy} that decouples global topology from 
high-frequency surface detail, enabling each stage to specialize 
on a well-defined subproblem.

\textbf{Stage 1: Coarse Structure Generation.} 
The first stage establishes a geometric foundation by generating 
coarse latents directly from image conditioning, using a scaled-up  Seed3D 1.0 DiT backbone. While this stage reliably 
captures overall topology and coarse surface structure, the 
resulting shapes tend to exhibit over-smoothed geometry with 
degraded high-frequency detail, as sharp edges, precise curvatures, 
and fine surface features are not faithfully recovered. We attribute 
this to the inherent tension between learning global structure and 
local detail simultaneously within a single unguided diffusion 
process.

\textbf{Stage 2: Detail Refinement.} 
{The second stage takes the Stage 1 output as a geometric anchor and refines the latents to recover high-frequency geometric detail that single-stage generation fails to capture. Two complementary priors regularize this process:}
\begin{itemize}
    \item \textbf{Coarse Shape Prior.} 
    {The Stage 1 latents 
    are partially diffused and incorporated into the Stage 2 diffusion process as a coarse geometric reference, guiding refinement toward local detail recovery rather than global structure generation.}
    \item \textbf{Voxelized Positional Encoding.} 
    {The coarse geometry 
produced in Stage 1 provides voxelized spatial coordinates that 
are injected into the Stage 2 diffusion process as positional 
encodings, anchoring each latent token to an spatial location and thereby promoting structural regularity in the refined output.}
\end{itemize}
{Conditioned on these priors, the Stage 2 DiT focuses exclusively 
on generating sharp edges and fine surface details, bringing the 
final output closer to the reconstruction fidelity achievable by 
VAE.
}

\subsection{Texture}

\begin{figure*}[t]
  \centering
  \includegraphics[width=\linewidth]{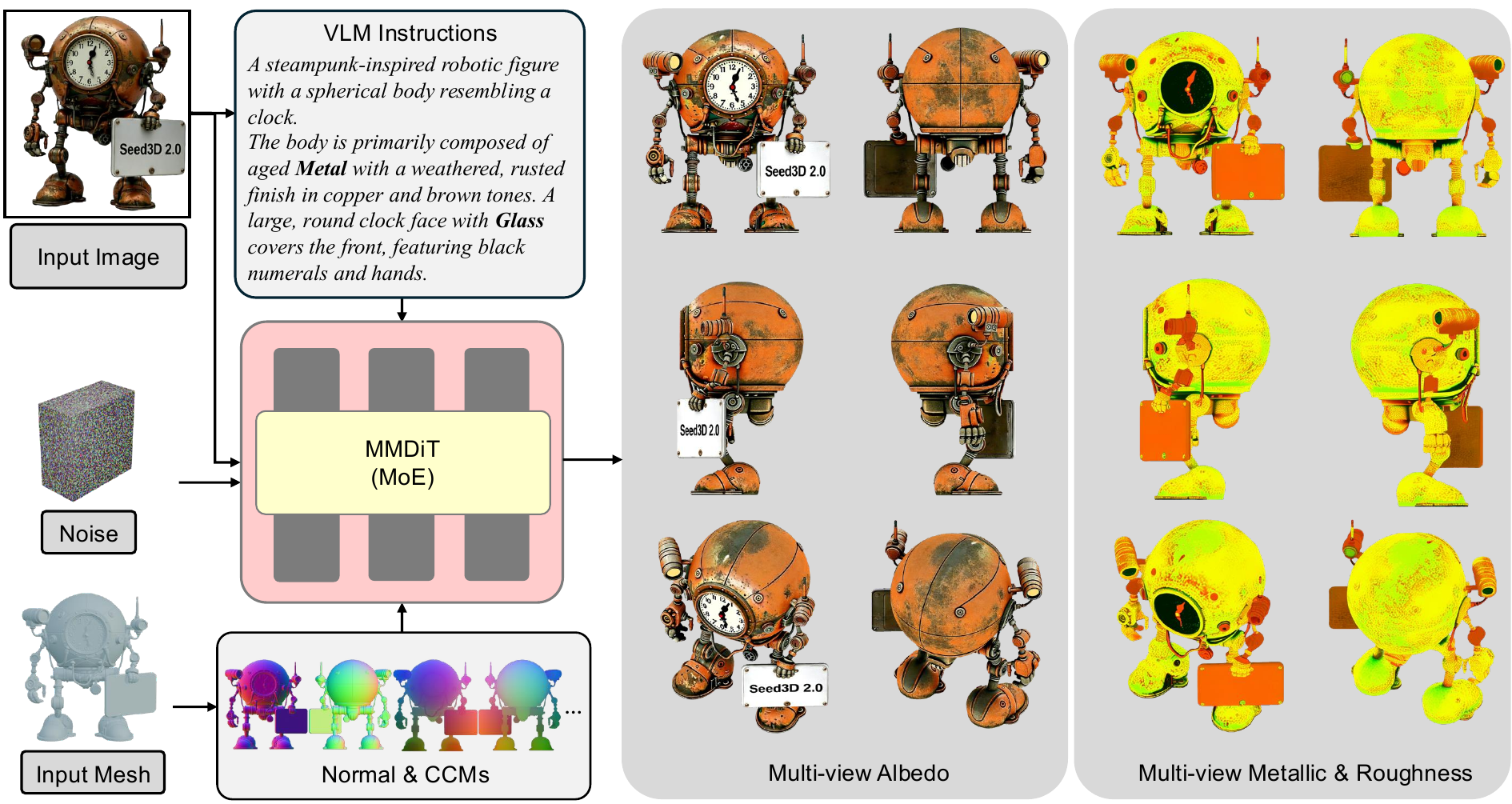}
  \caption{{Overview of the Seed3D 2.0 texture generation pipeline.} A unified model generates multi-view albeo and material under the guidance of VLM-based priors.}
  \label{fig:texture_pipeline}
\end{figure*}

Seed3D 2.0 Texture consolidates the cascaded texture pipeline shown in Seed3D 1.0~\cite{seed2025seed3d} (Seed3D-MV and Seed3D-PBR) into a unified PBR generation model that directly produces multi-view albedo and metallic-roughness (MR) maps from reference images and 3D geometry. This eliminates the intermediate multi-view 
RGB synthesis step and the associated error accumulation. As shown in Figure~\ref{fig:texture_pipeline}, the model retains the MMDiT-based two-stream architecture of Seed3D 1.0, jointly modeling albedo and MR through modality-specific projection layers within shared DiT blocks, and extends it with a Mixture-of-Experts (MoE) design and VLM-based semantic conditioning to improve resolution, efficiency, and generation robustness.

\textbf{Higher-Resolution {Texture} Generation.}
In Seed3D 1.0, the low image resolution limits the model's ability to preserve fine-grained visual details during {texture} decomposition. To improve {texture} fidelity, we increase the generation resolution of the unified PBR model. 
However, simply enlarging the latent space with a dense architecture would impose prohibitive computation cost. We therefore adopt a Mixture-of-Experts (MoE)~\cite{shazeer2017outrageously} design, expanding the network capacity while keeping less active computation through sparse expert routing. 
The higher resolution and stronger model capacity enable richer visual patterns, more detailed surface textures, and improved fidelity of fine-grained features such as text and patterns than Seed3D 1.0, particularly benefiting albedo generation quality. The precision of metallic-roughness boundaries is also improved, producing cleaner {texture} transitions across surface regions.

\textbf{VLM Prior Conditioning.} {PBR estimation under unknown illumination is inherently ill-posed: identical appearances can arise from different combinations of lighting and {material} properties. Common failure modes include color shifts where lighting effects are incorrectly baked into {albedo}, specular highlights on {non-metal} surfaces misclassified as metallic responses, and {metallic} surfaces wrongly estimated as {non-metals} under diffuse illumination. Such ambiguities fundamentally limit the stability and physical plausibility of purely appearance-based {material} estimation.} To address this, we introduce a Vision-Language Model (VLM)~\cite{seed16} prior as an additional conditioning signal. VLM-generated descriptions of {material} types, surface characteristics, and physical attributes are encoded into conditioning tokens and injected into the DiT blocks together with the geometry and reference image conditions, providing semantic grounding that stabilizes {material} {generation} and improves physical plausibility.

\subsection{Simulation-Ready Generation Model Suites}\label{sec:simulation-ready-models-suites}

Building on the high-fidelity geometry and texture 
generation capabilities of Seed3D 2.0, we further extend 
the framework with a suite of simulation-ready generation 
models that transform image-to-3D outputs into assets that 
can be placed, decomposed, and articulated for downstream 
applications. The overall system follows a staged workflow 
proceeding from scene-level layout planning to 
instance-level part-aware decomposition and kinematic 
modeling, comprising three modules:

\begin{itemize}
    \item \textbf{Scene Layout Planning.} Predicts spatially 
    consistent object layouts from text, image, or video 
    inputs, and composites the generated assets into 
    coherent 3D scenes, extending instance-level generation 
    to scene-level compositional asset creation.

    \item \textbf{Functional Part Decomposition.} Partitions an object into semantically and functionally coherent components, providing the structural basis required for physical interaction.

    \item \textbf{Articulation Generation.} Infers the kinematic structure of decomposed parts, including part hierarchy, joint types, axes, and motion ranges.
\end{itemize}

\subsubsection{Scene Layout Planning}\label{sec:simulation-ready-scene}

Extending object-level generation to full-scene synthesis requires inferring where objects should be placed, how they should be scaled, and how they relate to one another within a coherent 3D environment.
As the available geometric and spatial cues differ fundamentally between visual and textual inputs, we design two strategies tailored to each modality.

{For visual inputs, especially a single input video, the system must recover global scene structure, spatial arrangement, and per-object appearance from partial, view-dependent observations.}
We first leverage depth estimation to recover scene geometry and infer the global distribution of objects over time. Frame-wise object detection and segmentation yield precise instance masks, which are then passed to a VLM to generate textual descriptions for each entity; image inpainting is applied to recover occluded regions and produce a more complete visual profile per object.
The refined object images are fed into Seed3D 2.0 geometry and texture generation models to generate high-fidelity meshes, which are subsequently aligned with the estimated depth map to determine the spatial coordinates and scale of each object, producing a coherent scene layout for compositon.

For text inputs, scene layout planning is inherently underconstrained:
the system must infer both scene composition and spatial relationships among objects from language alone, without explicit geometric or positional cues.
To address this, we fine-tune an LLM for spatial reasoning, enabling it to generate plausible object layouts and per-object descriptions from text alone.
Seed3D 2.0 geometry and texture models are then conditioned on each object's layout and description to generate individual assets, which are composited into a unified scene.

\subsubsection{Part-level Generation}\label{sec:partgen}
\begin{figure}[t]
    \centering
    \includegraphics[width=0.9\linewidth]{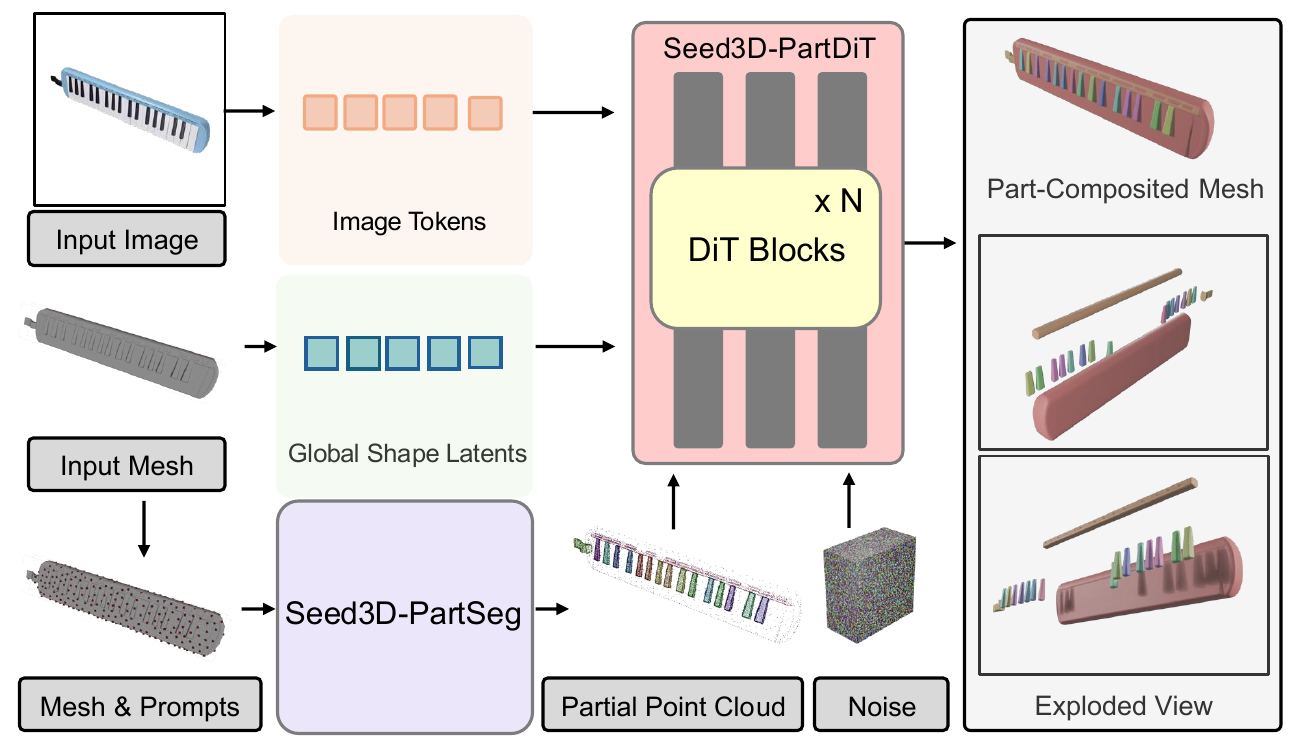}
    \caption{{Overview of the part-level generation pipeline.} A given 3D mesh will be first segmented to obtain partial point cloud by Seed3D-PartSeg, and then passed to Seed3D-PartDiT to get the final composited mesh.}
    \label{fig:partgen_pipeline}
\end{figure}

{For simulation-ready assets, whole-object geometry alone is insufficient for fine-grained physical interaction; many manipulation tasks require understanding and controlling individual functional components such as doors, drawers, or handles. A key challenge is therefore parsing generated shapes into functionally meaningful parts that can be independently controlled and simulated.} {To achieve part-level generation,} we adopt a two-stage pipeline following a ``perception-then-generation'' paradigm: (i) Seed3D-PartSeg, a part-level segmentation model, that decomposes the asset into coherent surface regions; and (ii) Seed3D-PartDiT, a part-completion model that reconstructs the corresponding part-aware 3D asset. The overall pipeline is illustrated in Figure~\ref{fig:partgen_pipeline}.

Seed3D-PartSeg takes a generated mesh as input and produces surface segments corresponding to functional components. 

{
    A native 3D backbone~\cite{sonata,Wu2023PointTV} first extracts geometric features from points randomly sampled from the mesh surface. The features are then processed by segmentation heads to produce part masks conditioned on sparse point prompts. Following non-maximum suppression (NMS) filtering, the point-level predictions are projected onto mesh faces and propagated to unlabeled regions to yield a complete surface segmentation.
}

{Building on these segments, Seed3D-PartDiT generates each decomposed part using a rectified flow-based diffusion framework, conditioned on three complementary signals: global shape latents extracted from Seed3D-VAE for structural context, partial point clouds from Seed3D-PartSeg for spatial guidance, and input image for appearance 
reference.
A modified attention design enforces both inter-part and intra-part interactions during denoising~\cite{Lin2025PartCrafterS3}, while global shape features are additionally injected into each DiT block to further reinforce geometric consistency across parts.
Each denoised latent is decoded by the Seed3D-VAE decoder and assembled into a part-composited mesh.}

\subsubsection{Articulated Asset Generation}\label{sec:articulated-gen}

{While part decomposition recovers the compositional structure of an object, it does not characterize the kinematic relationships among components, which are essential for any downstream application requiring physically plausible part-level interaction.
Large-scale articulated 3D supervision is scarce and expensive to obtain; we therefore develop a training-free pipeline that infers articulation structure by combining three complementary sources of prior knowledge: semantic priors from Vision-Language Models (VLMs), geometric priors from the decomposed mesh, and dynamic priors from image-to-video generation models.
}

We first employ VLMs on rendered views to organize decomposed components into kinematically coherent parts and identify their joint types. 
For each articulated part, we generate a pool of joint axis candidates based on part geometry using predefined geometric operators, and prompt a VLM to select the most plausible candidate, combining the precision of geometry-based candidate generation with the semantic reasoning of VLMs for adjudication. 
For motion range estimation, static geometry alone is 
inherently underconstrained: the same part can admit a wide 
spectrum of physically valid motions. We resolve this 
ambiguity by leveraging an image-to-video generation model 
as a motion prior, exploiting motion knowledge learned from large-scale natural videos~\cite{li2024dragapart,song2025puppeteer}.
Specifically, given a rendered image and a 
text prompt describing the intended actuation, the model 
synthesizes a short clip of the part in motion, from which 
we fit the joint range via differentiable rendering~\cite{pytorch3d} against the generated motion sequence, following prior motion reconstruction works~\cite{liu2023paris,song2025puppeteer}.

The resulting articulation structure, {together with basic physical properties (e.g., mass and friction) estimated by a VLM,} is exported to standard formats such as URDF, enabling the generated assets to be deployed in physics and graphics engines for downstream application requiring physically plausible part-level interaction.

\section{Data}
\label{sec:data_preprocessing}

\begin{figure*}[t]
  \centering
  \includegraphics[width=\linewidth]{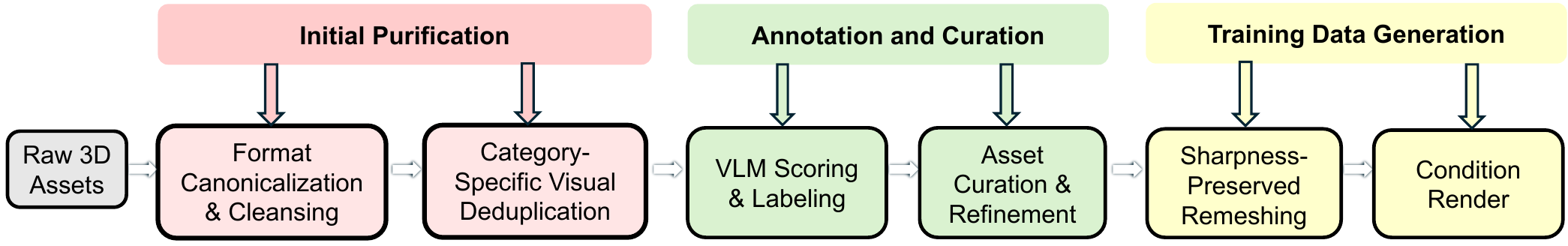}
  \caption{{Overview of Seed3D 2.0 data pre-processing framework.} We design a six-stage data pre-processing pipeline to ensure the quality of training datasets.}
  \label{fig:data_pipeline}
\end{figure*}

As illustrated in Figure~\ref{fig:data_pipeline}, the processing pipeline for Seed3D 2.0 consists of six core stages which ensures the quality and diversity of our training dataset.

\textbf{Format Canonicalization and Cleansing.} Raw assets are first converted into a unified representation coupling mesh geometry with multi-channel PBR textures. This is followed by geometric purification to remove ``pseudo-3D'' billboard artifacts and extraneous structures (\textit{e.g.}, pedestals), and texture verification to exclude assets with corrupted UV layouts or missing texture channels.

\textbf{Category-Specific Visual Deduplication.} To balance data diversity and uniqueness, we apply category-specific dynamic thresholds to 2D features extracted from multi-view renders, preventing over-filtering in visually homogeneous categories while aggressively eliminating near-duplicates.

\textbf{Advanced VLM Scoring and Captioning.} 
{We fine-tuned VLMs to evaluate each asset across six dimensions spanning semantic, structural, and perceptual criteria, with a powerful LLM serving as arbiter.}
Standardized text captions are also generated per asset. These multi-dimensional labels and captions serve as the foundational criteria for downstream curation and as semantic priors for model conditioning.

\textbf{Asset Curation and Refinement.} 
Assets are branched into pre-training and supervised fine-tuning (SFT) subsets. VLM tags are used to filter low-quality candidates and guide targeted refinement, including canonical orientation alignment to standardize object poses and instance disentanglement to resolve semantically ambiguous multi-object layouts. The SFT subset undergoes additional fine-grained human verification for geometry and texture fidelity.

\textbf{Sharpness-Preserved Watertight Remeshing.} Curated meshes are converted into a DMC-compliant watertight representation via a fully GPU-accelerated pipeline. 
Rather than conventional $L_2$ distance minimization, we adopt a sharpness-preserving formulation inspired by the $L_{\infty}$ metric to maintain dihedral angles and edge discontinuities.
The optimized CUDA pipeline achieves $1024^3$ resolution reconstruction within 15 seconds. 
The same remeshing process is applied independently to each part if the 3D asset consists of multiple parts.

\textbf{Condition Rendering.} In the final stage, we generate high-quality conditioning signals for the diffusion process, including view-consistent geometric rendering along uniform camera trajectories, and multi-channel PBR texture rendering of albedo, roughness, and metallic maps.

\section{Model Training}

\subsection{Geometry}


Seed3D-DiT training follows a hierarchical progressive pipeline, scaling from foundational structure learning to high-frequency detail refinement across two stages.

\textbf{Stage 1: Foundational Training.}
The first stage establishes a robust mapping from image conditioning to 3D latent representations through three successive phases.
\begin{itemize}
    \item \textbf{Pre-Training (PT).} The model is trained from scratch on a broad-scale dataset at a base resolution of 256 latent tokens and 256-resolution images, learning the foundational distribution of 3D shapres and initial cross-modal alignment betwen visual features and geometric representations.
    \item \textbf{Continued Training (CT).} To capture more intricate surface structures, we progressively increase the latent sequence length to 4096 and image resolution to 512, enabling the model to learn finer geometric details and sharper edge preservation.
    \item \textbf{Supervised Fine-Tuning (SFT).} Following the large-scale training, we perform fine-tuning on a high-quality, curated subset with reduced learning rates to eliminate unwanted surface perturbations and improve overall surface topology.
\end{itemize}

\textbf{Stage 2: Precision Refinement Training.}
{The second stage specializes in recovering high-frequency geometric detail that single-stage generation fails to capture, using stronger geometric regularization and curated data.}
\begin{itemize}
    \item \textbf{Initialization.} The Stage 2 DiT is initialized from Stage 1 checkpoints, allowing the model to leverage existing structural knowledge as a starting point for refinement.
    \item \textbf{Continued Training (CT) with Regularization.} 
    {Building upon the Stage 1 foundation, we conduct continued training incorporating Voxelized Positional Encoding that provide spatial constraints, and partially diffused Stage 1 latents that serve as a coarse geometric anchor for details recovery.}
    \item \textbf{Advanced SFT for High Quality Generation.} Following CT, we finetune the model on a carefully curated set of high-quality samples, targeting improved sharpness, geometric regularity, and fidelity to input reference images.
\end{itemize}

\subsection{Texture}

Seed3D-PBR training adopts a two-stage progressive training strategy. The first stage establishes broad {texture} generation capabilities, while the second stage refines quality through semantic guidance.

\textbf{Pre-Training.}
{
    The unified PBR model with MoE architecture is trained on a large-scale dataset, learning foundational capabilities in multi-view albedo and metallic-roughness generation conditioned on reference images and 3D geometry. This phase maximizes coverage of diverse albedo appearances, {metallic-roughness} categories, and lighting conditions to establish broad {texture} generation ability before quality-focused refinement.
}

\textbf{Supervised Fine-Tuning (SFT).}
{
    Following pre-training, we finetune on a carefully curated high-quality subset with reduced learning rates. VLM-generated {material} descriptions are introduced at this stage as an additional conditioning signal, enabling semantically-guided material estimation. By deferring VLM integration to SFT, the model first consolidates general {texture} generation capabilities during pre-training before specializing in resolving ambiguous {material} decomposition under complex lighting conditions.
}
\section{Inference}

\subsection{Inference Pipeline}
Inheriting the multi-stage inference paradigm of Seed3D 1.0~\cite{seed2025seed3d}, our system generates high-fidelity textured 3D assets through a sequential pipeline including geometry generation, {texture} synthesis, and UV texture completion.

\noindent\textbf{Geometry Generation.}
Given an input image, the Stage~1 DiT first predicts a coarse VecSet latent, which is decoded into an intermediate mesh via DMC on a sparse $512^3$ grid. This coarse mesh is then re-encoded into latents via the VAE encoder; in parallel, it is voxelized through GPU-accelerated voxelization and morphological dilation to produce a spatial occupancy prior. Together, these two signals condition the second stage for high-resolution mesh generation.
To support extraction at resolutions up to $1536^3$, we employ a hierarchical strategy that progressively prunes spatial query points using Stage~1 occupancy prior and multi-scale filtering, while improving efficiency through spatially-aware grouping in cross-attention-based SDF querying. Finally, the mesh is simplified using GPU-accelerated QEM decimation~\cite{qem,trellis2} to the target face count, followed by UV unwrapping for {texture} generation.

\noindent\textbf{Texture Generation.}
The texture inference pipeline extends Seed3D 1.0 with parallized model execution and optimized post-processing operators, reducing end-to-end latentcy while maintaining generation quality.

\subsection{Inference Efficiency}
To reduce inference cost across all DiT models in Seed3D 2.0 for both geometry and {texture} generation, we adopt a two-stage progressive distillation pipeline.

In the first stage, we distill the classifier-free guidance (CFG) mechanism by training a student model to predict the CFG-combined output in a single forward pass, halving the per-step computation while preserving the quality benefits of guided sampling.
In the second stage, we apply progressive step distillation following the curriculum strategy introduced in~\cite{salimans2022progressive}, iteratively halve the sampling steps: at each round, the student learns to match its teacher's two-step output in a single step.
This staged compression avoids the training instability of single-stage aggressive distillation and yields a more faithful approximation of the original trajectory.

The resulting distilled model achieves performance comparable to the full-step CFG-guided model across three dimensions: visual fidelity, multi-view consistency and reference image alignment.

\section{Model Performance}
We present a comprehensive evaluation of Seed3D 2.0 under two settings:
(1) 3D shape-only generation and (2) textured 3D asset generation.
\begin{figure*}[!h]
  \centering
  \includegraphics[width=0.91\linewidth]{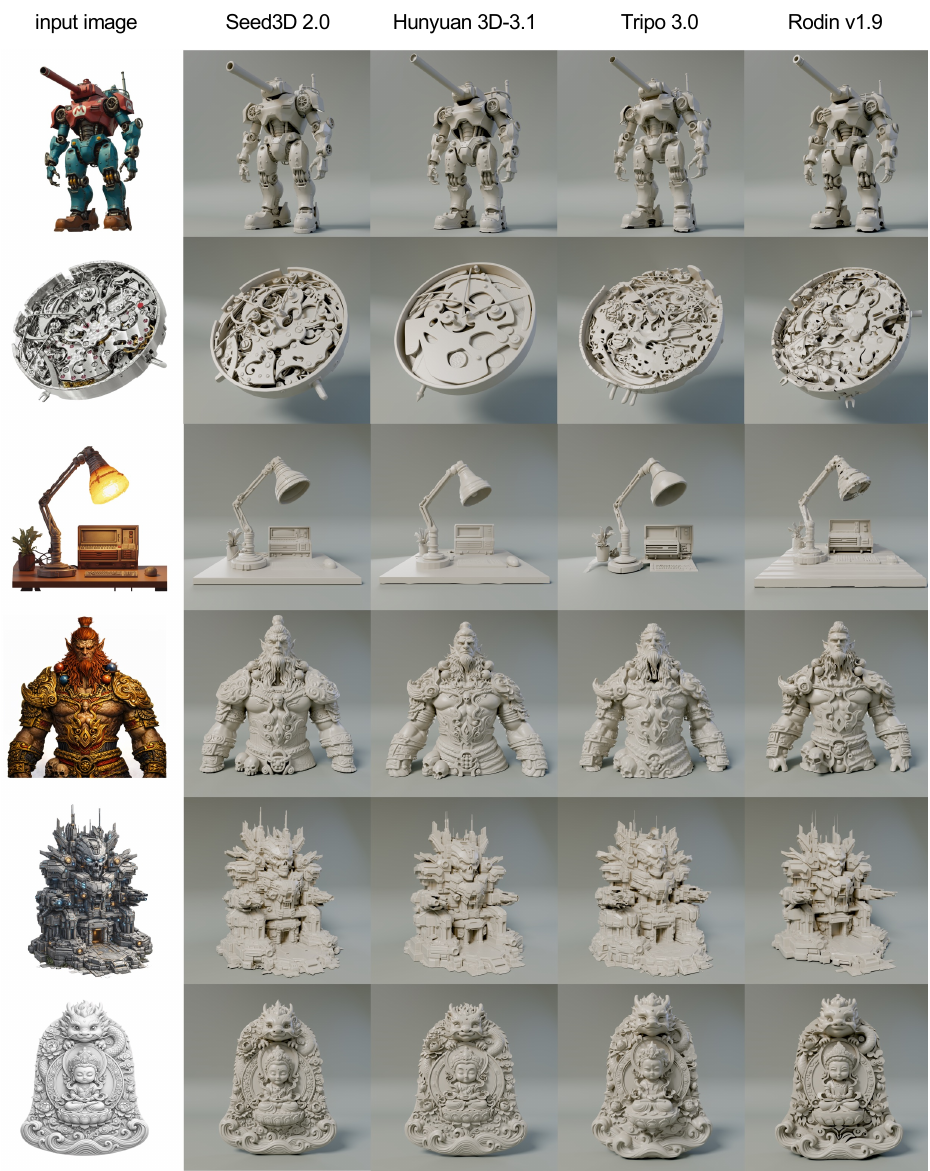}
  \caption{{Qualitative comparisons between Seed3D 2.0 and baselines in terms of 3D shape generation.}}
  \label{fig:geometry_comparison}
\end{figure*}

\begin{figure*}[!h]
  \centering
  \includegraphics[width=0.91\linewidth]{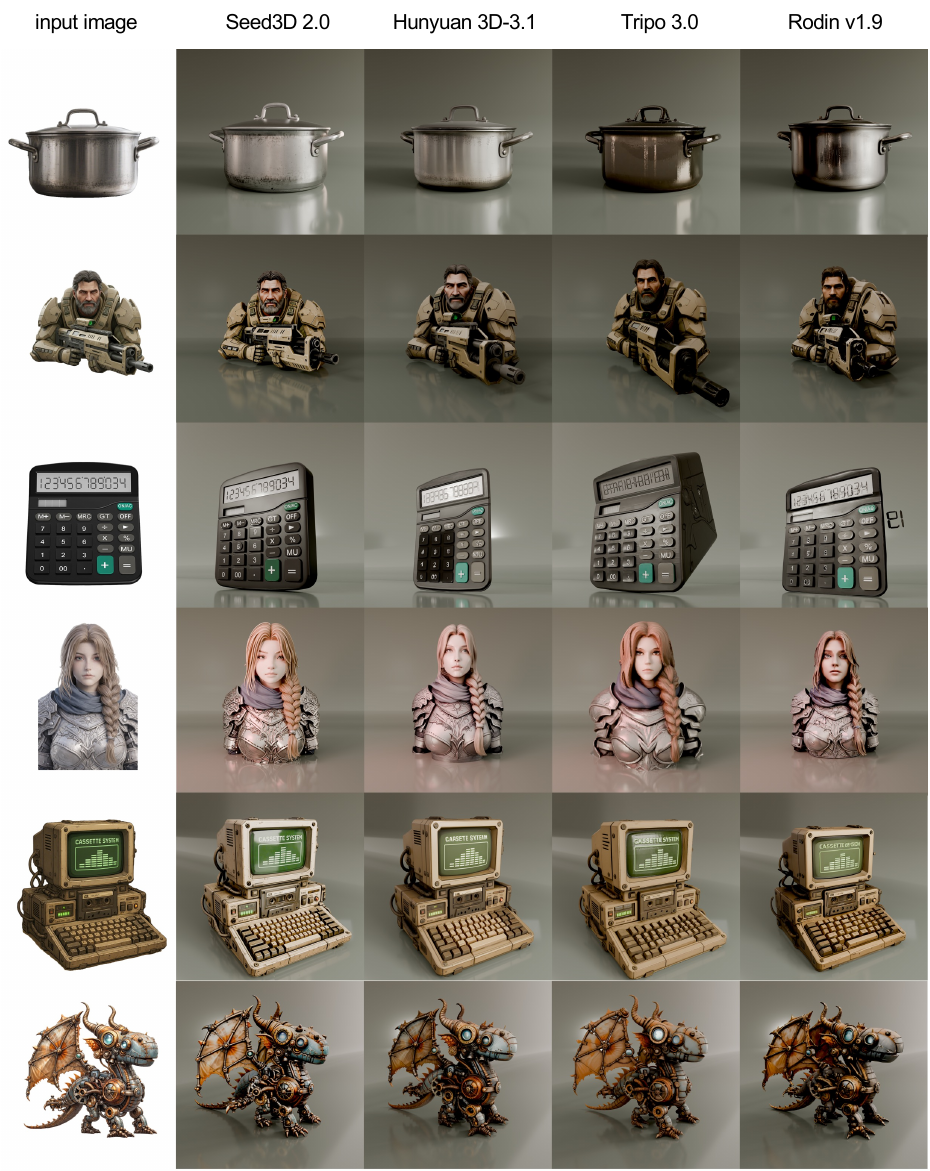}
  \caption{{Qualitative comparisons between Seed3D 2.0 and baselines in terms of {textured 3D asset generation.}}}
  \label{fig:texture_comparison}
\end{figure*}

We first conduct a user study to assess perceptual quality of generated 3D assets through blind paired comparisons against five recent methods: Hunyuan3D-2.5, Hunyuan3D-3.1, Tripo 3.0, Rodin Gen2 v1.9, and HiTem v2.0. Specifically, we recruited 60 human raters with 3D modeling background, randomly assigning 15 raters to evalute over 200 image prompts for each comparison. For each pair, raters are asked to judge which result is better, or whether the two are comparable. Figure~\ref{fig:quantitative_results} reports human preference results under two comparison settings, \textit{i.e.}, shape only and end-to-end textured assets.

For shape-only generation, Seed3D 2.0 consistently outperforms all compared methods. Win rates ranging from 55.2\% against Hunyuan3D-3.1, to 98.3\% against Seed3D 1.0, with clear margins against Tripo 3.0 (92.8\%), Rodin Gen2 v1.9 (89.6\%) and HiTem v2.0 (79.2\%). The near-unanimous preference over Seed3D 1.0 underscores the substantial geometric advances brought by the coarse-to-fine two-stage DiT and locality-aware VAE design in Seed3D 2.0.

For end-to-end textured asset generation, Seed3D 2.0 is preferred in the majority of cases across all compared methods, with win rates ranging from 69.0\% against Hunyuan3D-3.1 to 89.9\% against Rodin Gen2 v1.9. The consistently small proportion of comparable judgments suggests that quality differences are generally perceptible to human raters.

The results of user study can be further verified by qualitative comparisons. Figure~\ref{fig:geometry_comparison} demonstrates that Seed3D 2.0 produces high-precision geometry with sharp structural detail and strong faithfulness to the input image. 
Figure~\ref{fig:texture_comparison} shows clear advantages in {texture} quality and visual fidelity, including accurate {texture} decomposition and reliable text rendering.

\section{Application}
\subsection{Object-Compositional Scene Generation}

Many practical downstream applications require coherent 
multi-object environments rather than isolated assets, 
including synthetic scenes for embodied AI and XR content creation, where spatial and functional consistency across objects is as important as individual asset fidelity. Built on the instance-level 
generation quality of Seed3D 2.0 and the simulation-ready 
model suite in Section~\ref{sec:simulation-ready-models-suites}, 
we extend the framework to object-compositional scene 
generation. Given a user prompt (image/video or text), the 
scene-layout planning module in Section~\ref{sec:simulation-ready-scene} first constructs 
a 3D spatial layout; the instance-level pipeline then 
synthesizes and arranges the corresponding objects into a 
composited scene. This elevates 3D generation from asset creation to scene construction, providing a scalable foundation for any application that demands spatially coherent multi-object environments.

\subsection{Simulation-Ready Object Generation}\label{sec:app:sim-ready-object}
Beyond geometry and appearance, a simulation-ready asset must also expose functional structure and kinematic behavior, since interactive agents act upon meaningful parts subject to physical constraints rather than holistic meshes. We extend Seed3D 2.0 with two complementary capabilities: part-aware generation and articulated asset synthesis.

\begin{figure}[!h]
    \centering
    \includegraphics[width=\linewidth]{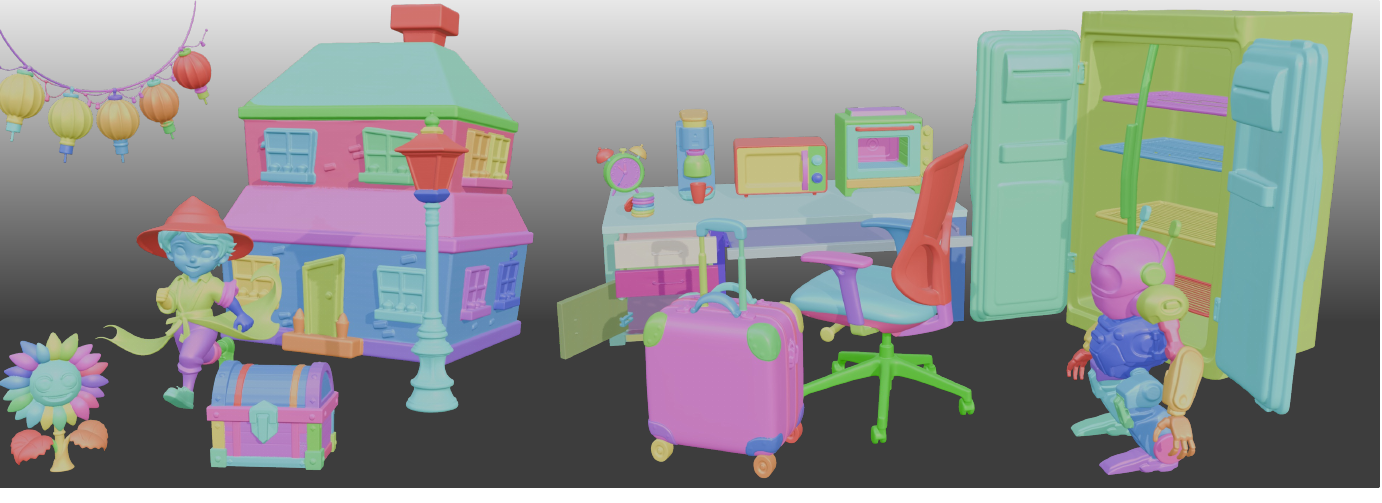}
    \vspace{-5mm}
    \caption{\text{Part-level generation capability of Seed3D 2.0.} Given a 3D mesh, the model decomposes the asset into functional parts for downstream applications.}
    \label{fig:partgen-demo}
\end{figure}

\noindent\textbf{Part-aware generation.} 
Seed3D 2.0 supports component-level modeling through the part-level pipeline described in Section~\ref{sec:partgen}. By applying Seed3D-PartSeg and Seed3D-PartDiT, we generate decomposed assets across a range of object categories, 
from architectural structures and 3D characters to 
household objects (Figure~\ref{fig:partgen-demo}).
The resulting decomposition exposes semantically meaningful 
units that can be directly reused for downstream articulation.

\begin{figure}[t]
    \centering
    \includegraphics[width=\linewidth]{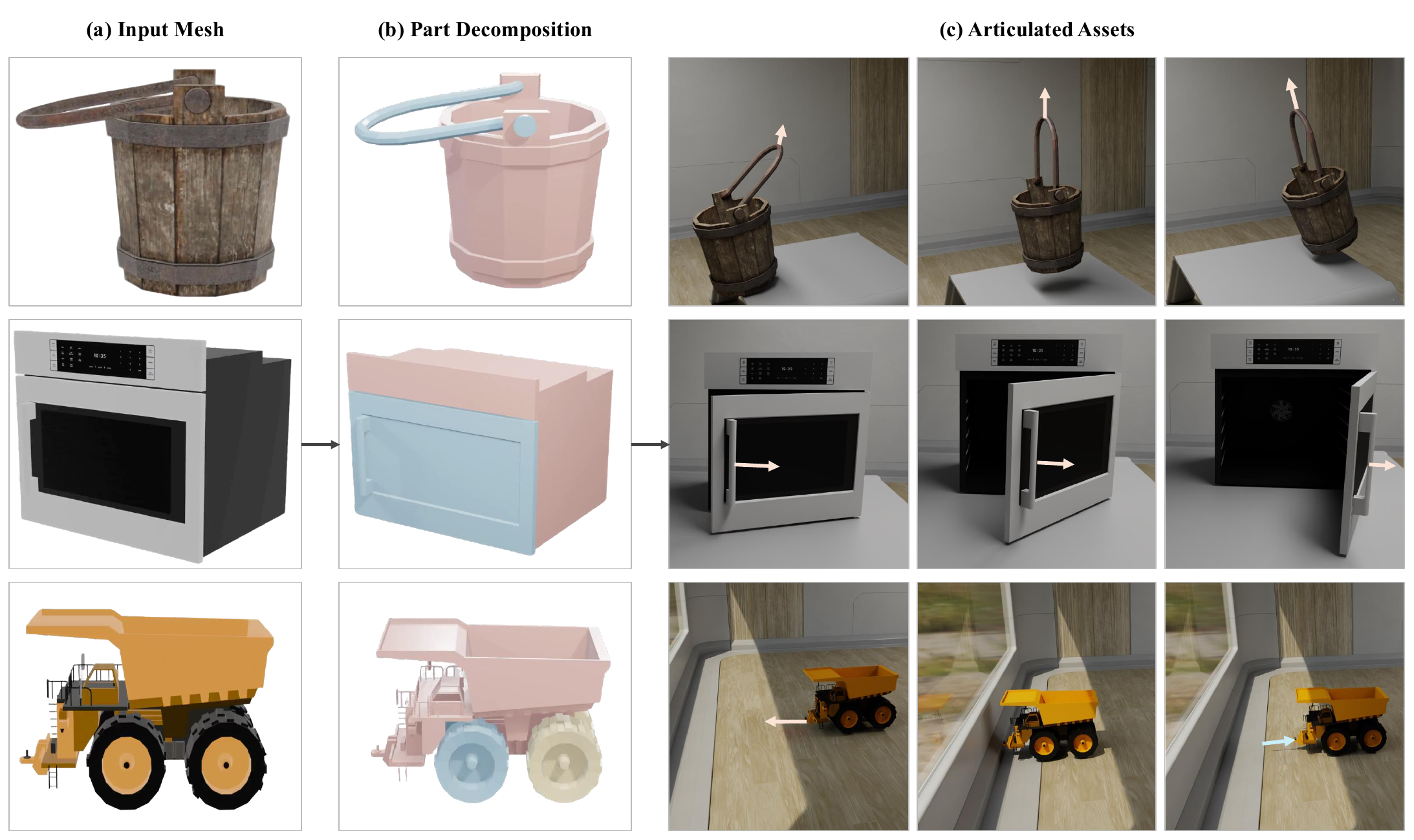}
    \vspace{-5mm}
    \caption{\text{Application to articulated asset generation.} Our framework enables the creation of interactive, simulation-ready assets from static 3D models. Given an input mesh, we perform part decomposition to parse the object into functional components, and then estimate articulation parameters to infer the kinematic structure and motion range. The resulting articulated assets support realistic part-level physical interactions, where the arrows indicate forces applied during simulation in Isaac-Sim~\cite{NVIDIA_Isaac_Sim}.}
    \label{fig:articulation-demo}
\end{figure}

\noindent\textbf{Articulation generation.} 
Building on the decomposed parts, the articulation pipeline introduced in Section~\ref{sec:articulated-gen} infers kinematic topology, joint types, axes, and motion ranges for each instance. {Combined with basic physical properties (e.g., mass and friction) estimated by a VLM, these attributes transform a static part-composited mesh into a fully articulated, simulation-ready asset exportable to standard formats such as URDF.} As shown in Figure~\ref{fig:articulation-demo}, the resulting assets can be integrated into scenes and driven by external forces in physics or graphics engines, enabling part-level physical interaction that extends beyond the capabilities of Seed3D 1.0~\cite{seed2025seed3d}.

\section{Conclusion}

We {present} Seed3D 2.0, a comprehensive system for 
high-fidelity, simulation-ready 3D content generation. 
On the geometry side, a coarse-to-fine two-stage generation 
pipeline with locality-aware VAE design substantially 
improves geometric sharpness and structural precision. For 
texture and material generation, a unified MoE-based PBR 
model with VLM semantic conditioning delivers improved 
material precision and visual fidelity. 
{Both advances are supported by a rigorous six-stage data 
curation pipeline that ensures high quality training data.}
These improvements are complemented by a simulation-ready 
model suite spanning scene layout planning, functional part decomposition, and articulated structure 
generation, bridging the gap between generative 3D modeling 
and real-world deployment requirements. A progressive 
distillation strategy further reduces inference cost across 
all generative models, supporting production-scale 
deployment.
Extensive user studies confirm the effectiveness of our 
approach: Seed3D 2.0 achieves consistent preference over 
all compared commercial models in both shape-only and 
end-to-end textured asset generation, with win rates of 
69.0\% to 89.9\% in end-to-end generation comparisons.
{Future works may extend several promising directions, such as richer physical attribute estimation, scaling scene generation to larger and more diverse environments, and deeper integration with downstream pipelines.}
We hope Seed3D 2.0 
serves as a meaningful step toward a unified, 
production-grade 3D content creation platform for both 
creative professionals and intelligent systems.

\newpage

\clearpage

\bibliographystyle{plainnat}
\bibliography{main}

\clearpage

\beginappendix
\section{Contributions and Acknowledgments}
\label{sec:contributions}

All contributors of Seed3D 2.0 are listed in alphabetical order by their last names.

\subsection{Core Contributors}
Diandian Gu, Jing Lin, Gaohong Liu, Jiahang Liu, Su Ma, Guang Shi, Jun Wang, Qinlong Wang, Qianyi Wu, Zhongcong Xu, Xuanyu Yi, Zihao Yu, Jianfeng Zhang, Zhuolin Zheng, Yifan Zhu

\subsection{Contributors}
Rui Chen, Hengkai Guo, Xiaoyang Guo, Mingcong Han, Xu Han, Xiu Li, Yixun Liang, Weiqiang Lou, Junzhe Lu, Guan Luo, Minghan Qin, Shuguang Wang, Yuang Wang

\subsection{Acknowledgments}
Weidong Chen, Yunpeng Chen, Jinxin Chi, Zixian Du, Luyao Guo, Liang Han, Lixin Huang, Kaihua Jiang, Shihao Jiang, Yuhan Li, Huan Liu, Ziwen Ma, Wanxing Wang, Xintuo Wang, Manlian Wu, Shilong Yang, Junyang Zhang, Eddie Zhou, Jiaqing Zhou, Tong Zhou

\end{document}